\documentclass[showpacs]{revtex4}
\usepackage{epsfig}
\newcommand{\beq}{\begin{eqnarray}}
\newcommand{\eeq}{\end{eqnarray}}
\begin{document}

\title{ Renormalization group and bound states 
\footnote[1]{
Presented at 48th Cracow School of 
Theoretical Physics: Aspects of Duality,
June 13-22, 2008, Zakopane, Poland} }
\author{ Stanis{\l}aw D. G{\l}azek } 
\affiliation{ Institute of Theoretical Physics, University of Warsaw }
\begin{abstract} 
Renormalization group procedure suggests that the
low-energy behavior of effective coupling constant 
in asymptotically free Hamiltonians is connected 
with the existence of bound states and depends on 
how the interactions responsible for the binding 
are included in the renormalization group equations. 
\end{abstract}
\maketitle

\section{Introduction}
\label{sec:I}

Using a very simple model with asymptotic freedom
(or limit cycle) and bound states, this lecture
explains a renormalization group (RG) procedure
for Hamiltonians, including the result that an
effective interaction strength grows at low
energies. Instead of integrating out high-energy
degrees of freedom, the procedure explained here
is carried out using similarity transformations.
The magnitude of increase of the interaction
strength (a coupling constant in an effective
Hamiltonian) at low energies is related in the
model to the existence of a bound state and a
degree to which the interaction responsible for
the effect of binding is included in the generator
of the similarity transformation. Most briefly,
the more interaction in the generator the less
increase in the coupling constant. In addition, 
since a tiny and quite generic modification of 
the simple model changes the asymptotically free 
behavior into a limit cycle of a large period, 
the model shows that an apparently asymptotically 
free behavior in a considered range of scales may 
actually be a part of a cycle with a period 
much larger than the considered range. In the 
limit cycle case, the model illustrates how a 
large set of bound states can influence behavior 
of the effective coupling constant at low energies, 
also depending on the generator used for the 
similarity transformation. Key references for 
this lecture are~\cite{SRGf} and~\cite{SDGeffective}. 

How the renormalization group findings concerning 
bound states may be related to the AdS/CFT 
correspondence~\cite{Maldacena,MaldacenaReport},
including the RG interpretation proposed by
Polyakov~\cite{PolyakovWall,stringsQCD}, is not 
yet established. On the other hand, the RG procedure
discussed here is particularly useful in the front 
form of Hamiltonian dynamics~\cite{Dirac1,Dirac2}.
This form of dynamics is precisely the one in 
which recent holographic models have been proposed 
for calculating masses and wave functions of bound 
states of quarks and 
gluons~\cite{BrodskyTeramond1,BrodskyTeramond2}.

Section~\ref{sec:Model} defines the model
Hamiltonian. Subsequent sections introduce
relevant concepts on the basis of elementary
calculations in the model. Ultraviolet divergences
are identified in Section~\ref{sec:UV}.
Section~\ref{sec:standardRG} reviews a standard RG
procedure. The review includes: derivation of the
result that a coupling constant increases at low
energies (without explicit discussion of
scattering processes or Green's functions); demonstration of
asymptotically free behavior of the coupling
constant; explanation of how counterterms are
calculated; and identification of difficulties
associated with the increase of the coupling
constant at low energies. Section~\ref{sec:SRG}
introduces the similarity RG (SRG) procedure. A
simple class of generators of similarity
transformations is introduced in
Section~\ref{subsec:Generators}. Solutions for
effective Hamiltonians in the SRG procedure with
different generators, are described in
Section~\ref{sec:Solutions}. The case of
asymptotic freedom is discussed in
Section~\ref{subsec:AFandf}, and the case of limit
cycle, related to asymptotic freedom, is discussed
in Section~\ref{subsec:AFandLC}.
Section~\ref{sec:C} concludes the lecture by a
summary of the connection found in the model
between the increase of effective interaction
strength at small energies and existence of bound
states.

A set of brief appendices is added in order to point 
out analogies between the simple model and quantum 
field theory (QFT), such as QCD. The appendices discuss: 
singular, $\delta$-function potentials in effective 
Schr\"odinger equations (Section~\ref{subsec:delta}), 
theory of effective particles that aims at explaining 
the origin of the constituent quark model and binding 
of partons (Section~\ref{subsec:SRGandQFT}), and some 
questions concerning a possible connection between 
the SRG procedure and AdS/CFT correspondence 
(Section~\ref{subsec:AdSCFT}).


\section{Model}
\label{sec:Model}

In order to understand the mechanism that causes
an asymptotically free interaction to rapidly
increase at small energies, a theoretical feature
often suggested to be associated with confinement,
we need a precise definition of what is meant by
the effective coupling constant. In theories as
complex as QCD, the concept of effective coupling
constant at low energies is not simple. Therefore,
it seems appropriate to first define the concept
using a simple model and see what happens there 
before one plunges into the complexity of QCD. 

The well-known asymptotically free coupling constant 
depends on the energy scale logarithmically (it is 
proportional to the inverse of the logarithm of the 
energy). The energy is actually a kinetic one, not 
involving interactions in any significant way, and
the scale should be conceptually associated with 
momentum, a kinematical variable of quantum mechanics 
rather than a dynamical one. Thus, when constructing 
a simple model Hamiltonian in the generic form of
\beq
H & = & H_0 + H_I \, ,
\eeq
one considers $H_0$ a free (kinetic) energy and $H_I$ 
an interaction. An alternative way of thinking is
that all eigenvalues and eigenstates of $H_0$ are
known and provide a basis for describing what
happens due to $H_I$. In both ways, for studies
of logarithmic effects of asymptotic freedom, it is 
useful to assume that the spectrum of $H_0$ is not 
degenerate and has the form
\begin{eqnarray}
H_0 |n \rangle & = & E_n |n \rangle 
\, , \quad \quad 
E_n  = \mu \, b^n  
\, , \quad \quad 
b>1 \, ,
\eeq
where $\mu$ is a unit of energy. For convenience,
$\mu$ is set to 1 and omitted from further
consideration (all numbers that refer to energy
need to be multiplied by $\mu$ in order to obtain
the actual quantity). The utility of using powers
of $b$ is that successive momentum scales, or
eigenvalues of $H_0$, are separated by a constant
on a logarithmic scale, the constant being $\ln
b$. Therefore, a small number of states, just one 
per momentum (or energy measured by $H_0$) scale, 
will be sufficient to track logarithmic effects
associated with asymptotic freedom and, at the
same time, control effects of binding. Once the
eigenstates of $H_0$ are normalized, $\langle
m | n \rangle = \delta_{mn}$, the matrix elements
of $H_0$ in the model are
\beq
\langle m | H_0 | n \rangle & = & E_m \delta_{mn} \, .
\eeq
The interaction Hamiltonian $H_I$ is defined in 
the model by its matrix elements, using assumption 
that these elements should be factorized, i.e.,
$H_{I mn} = H_m H_n$, for the purpose of having 
very simple exact solutions to the eigenvalue 
problem for $H$. Thus, the factor $H_n$ should 
have dimension of square root of energy. The 
simplest possibility is $H_n \sim \sqrt{E_n}$. The 
proportionality is reduced to a dimensionless
number, and one can write 
\beq
\label{HI}
\langle m | H_I | n \rangle   = 
- g \sqrt{E_m} \sqrt{E_n} \, ,
\eeq
where $g$ determines the strength of the
interaction. It is called coupling constant, in
analogy with the standard nomenclature in QFT. The
negative sign is chosen so that there exists a
bound state for a sufficiently large positive $g$.
By definition, the bound state corresponds to a
negative eigenvalue of $H$. 

In summary, matrix elements of the model Hamiltonian 
are
\beq
H_{mn} & = & 
\langle m | H | n \rangle 
=
E_m \delta_{mn} - g \sqrt{E_m} \sqrt{E_n}
\, .
\eeq
Note that it will have to be clarified what the 
words ``sufficiently large $g$'' are supposed to 
mean, because the model Hamiltonian produces 
divergences (infinities) no matter how small
the number $g$ is.

\section{ Ultraviolet divergences }
\label{sec:UV}

Consider first very small numbers $g$ and the intuition 
that eigenvalues of $H$ should be nearly equal to
the eigenvalues of $H_0$. For example, first-order 
perturbation theory produces a correction $\Delta E_m^{(1)}$ 
to the energy $E_m$ of the form
\beq
\Delta E_m^{(1)} 
& = & 
\langle m |H_I | m \rangle
  =
-g \, E_m \, .
\eeq
When $g$ is small, the correction is small; it is
just a fraction $g$ of the energy being corrected.
(This situation resembles what happens in the
Schr\"odinger equation for atoms or positronium
when one derives corrections to the Coulomb
potential from QED and calculates their influence
on energy levels in first-order perturbation
theory, see Appendix, Section~\ref{subsec:delta}.) 

Consider now the second-order correction to the
same energy 
\beq
\Delta E_m^{(2)} 
& = & 
\sum_{k \neq m} {  | \langle m |H_I|k \rangle |^2  \over E_m - E_k  } 
 =  
g^2 E_m \, \sum_{k \neq m} {E_k \over E_m - E_k} \, .
\label{2nd}
\eeq
Since $E_k = b^k$, a term number $k$ contributes
$1/(b^{m-k} - 1)$ in the sum. Thus, terms with $k
< m$ contribute a finite sum of a nearly geometric
series (with quotient $1/b$), which is not sensitive 
to any lower bound on $k$ in the sum, say $M$, if 
$b^M \ll b^m$. However, a lower bound is needed to 
define a Hamiltonian whose matrix has a finite size. 
In order to focus attention on $E_m \sim 1$, one 
can set the lower bound on $k$ to 
be a large negative integer $M$. As a result, all 
eigenvalues of $H_0$ included in the model are from 
now on assumed to satisfy the condition $E_k \ge b^M$, 
and $b^M \ll 1$. 

In contrast, terms with $k \gg m$ contribute each 
$-1$. For $b \gg 1$, the second-order correction is 
proportional to the number of basis states with 
energies (eigenvalues of $H_0$) greater than $E_m$. 
In order to obtain a finite answer, one must limit 
this number. For example, one can limit from above 
the range of summation over $k$ by certain large 
positive integer $N$. Then, one has to understand 
what happens when $N$ is very large. Imposing a limit 
such as $N$ is called ultraviolet regularization.

Including the ultraviolet regularization, the result 
of Eq. (\ref{2nd}) is approximated by the formula 
(the larger $b$, the better the approximation)
\beq
\label{N}
\Delta E_m^{(2)} 
& = & 
- g^2 E_m \, (N - m) \, .
\eeq
This formula shows that the correction tends to 
$- \infty$ when one attempts to send the
ultraviolet cutoff on energies, $\Lambda
= b^N$, to infinity. Since $N = \ln \Lambda/\ln
b$, the second-order correction is also approximately 
given by the formula
\beq
\label{Lambda}
\Delta E_m^{(2)} 
& = & 
- { g^2 E_m \over \ln b } \, \ln {\Lambda \over
E_m} \, .
\eeq
This result explains why the correction is called
logarithmically divergent in the ultraviolet. A
direct comparison of Eqs.~(\ref{N}) and
(\ref{Lambda}) illustrates that the ultraviolet
logarithmic divergence results from all different
energy scales contributing equally to the
eigenvalues. This conclusion is not limited to
perturbation theory. 

Consider the eigenvalue problem for the matrix 
$\left[H_{mn}\right]$,
\beq
\label{E}
\sum_{n = M}^N H_{mn} \psi_n & = & E \psi_m \, ,
\eeq
whose subscripts are limited after the regularization, 
$M \le m,n \le N$. Because the interaction is factorized, 
one has
\beq
\label{psi}
\psi_m & = & { g \sqrt{E_m} \, 
\over  E_m - E} \, \sum_{n = M}^N \sqrt{E_n} \psi_n 
\eeq
and the sum in Eq.~(\ref{psi}) is just a number (does 
not depend on $m$), say $c$. Substituting this solution 
for the wave function into Eq. (\ref{E}), one obtains
\beq
E_m { c g \sqrt{E_m} \over  E_m - E} 
- 
g \sqrt{E_m} \sum_{n = M}^N { c g E_n \over  E_n - E}
& = & 
E { c g \sqrt{E_m} \over  E_m - E} \, .
\eeq
or
\beq
1 + g \sum_{n = M}^N { E_n \over  E - E_n} & = & 0 \, .
\eeq
As far as the high-energy part is concerned, the sum in this 
non-perturbative eigenvalue condition is of the same type 
as in the perturbative Eq. (\ref{2nd}), the only change
being that $E_m$ with one particular value of $m$
is now replaced by the unknown eigenvalue $E$.
This does not change the fact that the sum can be
approximated by $ - \ln{ (\Lambda/|E|) }$. The result is 
that the eigenvalue condition has no meaning for finite $E$ 
when $\Lambda \rightarrow \infty$.

More specifically, consider a possibility that a
bound state exists, with a negative eigenvalue $E
= - E_B$. Replace the sum over $n$ by an integral 
with measure $dn = dE_n/(E_n \ln b )$, just to get an 
idea what happens. One has
\beq
\label{binding}
1 - { g \over \ln b }\int_{b^M}^\Lambda { dE \over  E_B +
E} & = & 0 \, ,
\eeq
or 
\beq
\label{E_B}
E_B & = & { \Lambda - b^M e^{\ln b \over g }
\over  e^{\ln b \over g } - 1 } \, .
\eeq
When one takes the limit of $\Lambda \rightarrow
\infty$ for fixed $g$, the binding energy
diverges linearly with $\Lambda$.

One can also observe that the square of the matrix
$H_I$ with large $N$ and $M$ is equal to ${g b
\Lambda \over 1 - b} \, H_I $, which means that it
diverges in the ultraviolet limit of $\Lambda
\rightarrow \infty$ for fixed $g$. This means that
all powers of the entire $H$ are ultraviolet
divergent. In particular, the evolution operator
$U(t,0) = e^{-iHt}$ does not exist in this limit. 

Note that the divergences in Eqs. (\ref{N}),
(\ref{Lambda}), and~(\ref{E_B}), result from the
diverging number of degrees of freedom. Similar
problems may occur in classical statistical
systems with a huge number of degrees of freedom
if all these degrees of freedom contribute
significantly to observables~\cite{WilsonRG}.

\section{ Standard renormalization group procedure }
\label{sec:standardRG}

Given the divergences, the task now is to figure
out whether to keep or discard the model that 
produces such diverging results when $N$ grows. 
Of course, one can keep $N$ fixed and try to 
describe physics (in the simple model, ``physics'' 
amounts to a set of eigenvalues and transition 
amplitudes). But if $N$ is expected (and hoped) 
to be very large (a desire a physicist has when 
designing a fundamental theory), the divergences 
can only be removed by making the coupling constant 
$g$ vanishingly small, and fine tuned to the value 
of $\Lambda = b^N$. On the other hand, a good model 
is expected to capture physics in a natural way, 
without worrying about very large numbers that by 
no means are comparable to the scale of observables 
of immediate interest. So, the trouble with the
diverging model can be summarized as follows.

Initially (in the model, by assumption; in
realistic theories, motivated by experience 
gathered through observations and experiments), 
one is led to believe that certain $H_I$ is a good 
candidate to consider. For example, in the model 
one may imagine that one measures transition 
rates between two states, say $|m_1 \rangle$ and 
$|m_2\rangle$, and that these rates can be 
described in first-order perturbation theory by 
an interaction Hamiltonian with matrix elements
\beq
\label{2x2}
\left[
\begin{array}{ll}
\langle m_2 | H_I | m_2 \rangle , & \langle m_2 | H_I | m_1 \rangle \\
\langle m_1 | H_I | m_2 \rangle , & \langle m_1 | H_I | m_1 \rangle \\ 
\end{array}
\right]
& = & 
-g
\left[
\begin{array}{ll}
E_{m_2}                  , & \sqrt{ E_{m_2} E_{m_1} }  \\
\sqrt{ E_{m_1} E_{m_2} } , & E_{m_1}                   \\ 
\end{array}
\right] \, . \nonumber \\
& & 
\eeq
One is then compelled to postulate that the whole
matrix of $H_I$ has the form given in
Eq.~(\ref{HI}). This way of thinking leads to 
divergences, as described in the previous section, 
and the question what to do about them. It is
clear now that one does not want to abandon the 
proposed interaction entirely since it does work 
for small $g$ in the cases of interest (in the 
example, for evolution of states built from $|m_1 
\rangle$ and $|m_2\rangle$) and exhibits appealing 
symmetry. In fact, this way of thinking is
used in building theories, by extrapolation from
known examples. An analogy in QFT is provided
by the gauge symmetry~\cite{YangMills}. Thus, 
one needs to conceive a general way out of the 
problem with divergences that are produced by 
naive extrapolation of knowledge from a small set 
of matrix elements to a large set of them. The 
large set is desired when one seeks a theory of 
presumably large range of applicability and a lot 
of predictive power.

The basic idea described here (using the model)
has been originally formulated in
Refs.~\cite{Wilson:1965,Wilson:1970}. The idea is
to learn what happens when one starts with some
large $N$ and tries to reduce the value of $N$ to
a smaller value, say $N_1$, and properly includes
all effects due to states between $N$ and $N_1$.
(This is sometimes called ``integrating out
high-energy degrees of freedom.'') The resulting
Hamiltonian $H^{(1)}$, limited in energy by $\Lambda_1
= b^{N_1}$, will contain information about what
has to be done in order to compensate for the
presence of the arbitrarily chosen cutoff
$\Lambda_1$. The step of reducing the size of a
cutoff can be repeated. One can reduce $\Lambda_1$ 
to $\Lambda_2$, $\Lambda_2$ to $\Lambda_3$, and so
on. Such steps are called RG transformations.
When a transformation is applied $K$ times,
a chain of Hamiltonians is obtained, including
$H$, $H^{(1)}$, ..., $H^{(K)}$. Eventually, two things
happen. 

First, the relationship between $H^{(K)}$ and
$H^{(K+1)}$ may become universal in the sense that
it no longer depends on all details of the initial
$H$. For example, if $H^{(K)}$ contains an
interaction term of the form $ - g^{(K)} \sqrt{E_m
E_n}$ and $H^{(K+1)}$ contains an interaction term
of the form $ - g^{(K+1)} \sqrt{E_m E_n} $, the
relationship between $g^{(K)}$ and $g^{(K+1)}$ may
be independent of the value of the initial
coupling constant $g$ in $H$. Instead, a universal
recursion is found for the coupling constant when
the cutoff is changed. (Such recursion may include
rescaling variables in order to compare successive
Hamiltonians in terms of functions of
dimensionless variables in a fixed range, leading
to anomalous dimensions.) We will see how the
recursion emerges in the model shortly. The model
illustrates this way how the existence of a
$\beta$-function in QED~\cite{GellMannLow} could
be understood, and explained in
Ref.~\cite{WilsonGML} in the context of strong
interactions. This is also how universality in
critical phenomena could be explained in classical
statistical mechanics~\cite{WilsonRG}. 

Second, after many RG steps, one obtains
Hamiltonians with running cutoffs $\Lambda_K$.
Suppose that one reduces the cutoff in every RG
step by a factor of $b$. Then $\Lambda_K =
b^{N-K}$ may be finite even when $\Lambda$ is sent
to infinity. What is required is that the number
of the RG steps, $K$, increases when one increases
$N$ for fixed $N-K$. Since the finite cutoff
$\Lambda_K$ can be chosen arbitrarily and, by
construction, eigenvalues smaller than $\Lambda_K$
for all $K$ considered do not depend on
$\Lambda_K$ at all, one eventually obtains a
family of effective Hamiltonians, $H_\lambda$ labeled 
by a finite cutoff parameter $\lambda = \Lambda_K$.
Predictions that follow from $H_\lambda$ do not
depend on $\lambda$.

\subsection{ Gaussian step }
\label{subsec:Gaussian}

It is time now to attempt the RG procedure
described in the previous section in the case
of our model. We will see how lowering $\lambda$ 
leads to increase of a coupling constant and why 
this increase causes trouble.

There are essentially two interrelated tasks to
accomplish. One is to establish the Hamiltonian
with the large cutoff $N$, so that the resulting
theory does not produce divergences (dependence of
observables on $N$). The other one is to evaluate
$H_\lambda$ with a finite $\lambda$ from the
well-defined initial $H$ with a priori
arbitrarily large $N$. The conceptual difficulty
of the RG procedure is that the first task is
accomplished in the process of trying to complete
the second one, and in the second task, the result
for $H_\lambda$ is used to decide how the initial
$H$ should be defined in order to make sure that
matrix elements of $H_\lambda$ do not depend on
$\Lambda$. The need for executing this process in
a sequence of successive approximations that can
converge on a structure that one is looking for,
is the hardest aspect of the procedure to
understand.

In the simple model, one can start with the eigenvalue 
problem 
\beq
H |\psi \rangle & = & E |\psi \rangle \, , \\
|\psi \rangle    & = & \sum_{k = M}^N \psi_k
|k\rangle \, ,
\eeq
in which the coefficients $\psi_k$ (a wave function) 
satisfy the set of linear equations~(\ref{E}).
One RG step is done by eliminating $\psi_N$ from
the remaining $N-M$ equations. Note that the initial 
number of equations is $N-M+1$, because there is one 
equation with $E_0 = 1$, in addition to $N$ equations 
corresponding to positive and $-M$ equations corresponding 
to negative powers of $b$. 

The eigenvalue problem is split into one equation 
for the highest energy component and a set of 
equations for the remaining components:
\beq
\label{first}
E_N \psi_N + \sum_{n=M}^N H_{I Nn} \psi_n & = & E \psi_N \, , \\ 
\label{rest}
E_m \psi_m + \sum_{n=M}^N H_{I mn} \psi_n & = & E
\psi_m \, , \quad M \le m \le N - 1 \, .
\eeq
Eq. (\ref{first}) produces
\beq
\psi_N = (E - E_N - H_{I NN})^{-1}
\sum_{n=M}^{N-1} H_{I Nn} \psi_k \, ,
\eeq
which is used in the remaining $N-M$ equations to
eliminate $\psi_N$ from them. In its essence, the RG
step is Gaussian elimination of one equation in a set 
of linear equations. The set involves the unknown 
eigenvalue $E$. The result of the first step is a set 
of equations with $M \le m \le N-1$,
\beq
E_m \psi_m + \sum_{n=M}^{N-1} H_{I mn} \, \psi_n +
\sum_{n=M}^{N-1}
{ H_{I mN} H_{I Nn} \over E - E_N - H_{I NN}} \, \psi_n
& = & E \psi_m \, .
\eeq
Therefore, the new interaction ``Hamiltonian'' in 
this one-step smaller set of equations has the 
following matrix elements ($M \le m,n \le N-1$):
\beq 
\label{H1}
H^{(1)}_{I mn}
& = & 
H_{I mn}
+
{ H_{I mN} H_{I Nn} \over E - E_N - H_{I NN}} \, .
\eeq
The word ``Hamiltonian'' is used in quotation
marks because the matrix elements of $H^{(1)}_I$
depend on the unknown eigenvalue $E$. Nevertheless,
Eq.~(\ref{H1}) is an exact result in the model. It 
guarantees that $H^{(1)} = H_0 + H^{(1)}_I$ with 
the cutoff $\Lambda_1 = b^{N-1}$ has the same 
eigenvalue $E$ as the Hamiltonian $H$ with 
the cutoff $\Lambda = b^N$ has. On the other 
hand, the operator $H^{(1)}_I$ is not fully 
defined before one specifies how to find the
eigenvalue it depends on. Some consistency
conditions would have to be imposed in a way
that still allows us finding the initial
Hamiltonian and reliably calculate $H_\lambda$. 

The situation simplifies considerably if one can
limit the RG procedure to eigenvalues $E$ that are
much smaller than a suitable finite cutoff $\lambda$ 
one wishes to reach. ``Suitable finite cutoffs 
$\lambda$'' in realistic theories are cutoffs that 
are small enough so that one can solve the eigenvalue 
problem for $H_\lambda$ on a computer. This condition 
puts severe constraints on the size of $\lambda$ in 
realistic theories, illustrated by, e.g., computational 
limitations of the lattice gauge theory. One has to 
execute $N-n$ relatively complex RG steps in order 
to reduce the cutoff from a formally infinite 
$\Lambda = b^N$ to some finite $\lambda$ 
and obtain $H_I^{(N-n)}$ in $H_\lambda$ with 
$\lambda = b^n$ that is sufficiently small for 
a reliable computation of the eigenvalues. 
$H_\lambda$ still includes a large number of matrix 
elements, on the order of $(n - M +1)^2$ times a
potentially large number of other degrees of freedom
besides the size of momentum, such as the numbers
of virtual particles and their angular momenta, 
spins, colors, or flavors.

In order to see how the simplification mentioned
above emerges for small eigenvalues $E$, and how
the simplification eventually ceases to be valid
in the case of asymptotically free Hamiltonians,
one may step back to Eq.~(\ref{H1}) and check what
happens in our model. Conclusions will not be
limited to the model case.

\subsection{ Asymptotic freedom in the model Hamiltonian }
\label{subsec:af}

In the model, Eq.~(\ref{H1}) reads
\beq 
H^{(1)}_{I mn}
& = & 
-g \sqrt{ E_m E_n }
+
{ (-g \sqrt{ E_m E_N }) \, (-g \sqrt{ E_N E_n }
\over E - E_N + g E_N} ) \\
& = & 
- \left( g - { g^2 E_N \over E - E_N + g E_N} \right)
\,
\sqrt{ E_m E_n } \, .
\label{H1model}
\eeq
It is clear that the interaction ``Hamiltonian'' 
$H_I^{(1)}$ has the same structure of matrix
elements as $H_I$ but contains a new ``coupling
constant,'' say $g^{(1)}$, that depends on the 
eigenvalue $E$. $g^{(1)}$ is given by the expression 
in the bracket in Eq.~(\ref{H1model}). 

The simplification for cutoffs much larger than 
$E$ becomes obvious when one re-writes
Eq.~(\ref{H1model}) as
\beq 
g^{(1)}
& = & 
g - { g^2 E_N \over E - E_N + g E_N} 
  =
g \, { 1 - E/E_N \over 1 - g - E/E_N } \, .
\label{g1model}
\eeq
One can neglect the ratio $E/E_N$ provided that the 
eigenvalue $E$ is small in comparison to the energy 
$E_N = b^N$. With this simplification, Eq.~(\ref{g1model}) 
reads
\beq 
g^{(1)}
& = & 
{ g \over 1 - g } \, ,
\eeq
and implies the following recursion in further RG steps 
\beq 
g^{(K)}
& = & 
{ g^{(K-1)} \over 1 - g^{(K-1)} } \, ,
\label{recursion}
\eeq
for as long as $E_{N-K} \gg |E|$ and $1-g \gg |E|/E_{N-K}$. 
The recursion of Eq.~(\ref{recursion}) is solved by
\beq 
g^{(K)}
& = & 
{ g \over 1 - g K } \, .
\label{solution1}
\eeq
Suppose one can solve the eigenvalue problem for
$H^{(K)} = H_{\lambda_0}$ with $\lambda_0 = b^{N-K}$ and 
establish that the coupling constant $g(\lambda_0) =
g^{(K)}$ should have some value $g_0$ in order to
reproduce some measured eigenvalue $E_0 \ll \lambda_0$. 
Eq.~(\ref{solution1}) says that 
\beq 
g_0
& = & 
{ g_\Lambda \over 1 - {g_\Lambda \over \ln b} \ln
\Lambda/\lambda_0 } \, .
\label{solution2}
\eeq
Eq.~(\ref{solution2}) allows one to calculate $g_\Lambda$ 
that needs to stand in the initial $H$ with the cutoff 
$\Lambda$ to produce the same eigenvalue. The
result is
\beq 
g_\Lambda
& = & 
{ g_0 \over 1 + {g_0 \over \ln b} \ln \Lambda/\lambda_0 } \, ,
\label{af}
\eeq
This means that the model is asymptotically free:
the larger the cutoff $\Lambda$ in the initial
$H$, the smaller the coupling constant $g_\Lambda$ 
in it. The most famous analogy in QFT is found in  
Refs.~\cite{af1,af2}.

\subsection{ Calculation of counterterms }
\label{subsec:CT}

Since Eq.~(\ref{af}) has been obtained from an exact
RG procedure under the two conditions that $|E_0|
\ll \lambda_0$ and $1-g_0 \gg |E_0|/\lambda_0$, 
one may expect that the replacement of $g$ by 
$g_\Lambda$ given by Eq.~(\ref{af}) provides a 
model $H$ with all eigenvalues $E$ that satisfy the 
same two conditions being also independent of 
$\Lambda$. Thus, these eigenvalues are also
expected to be common for all Hamiltonians 
$H_\lambda$ with $\lambda \ge \lambda_0$, i.e.,
they do not depend on the finite running cutoff 
$\lambda$ that limits the momentum range in 
$H_\lambda$.

The model calculation presented so far provides an 
example of how one takes the first step in identifying 
the initial $H$. The point is that the matrix elements 
of the initial $H$ can be now written as
\beq
\label{Hinitial}
H_{mn} & = & 
E_m \delta_{mn} - g \sqrt{E_m} \sqrt{E_n} +
CT_{mn} \, , \\
\label{CT}
CT_{mn} 
& = & 
(g - g_\Lambda) \sqrt{E_m} \sqrt{E_n} \, ,
\eeq
where the letters $CT$ are chosen as an abbreviation 
for the word ``counterterm.'' The counterterm
guarantees that small eigenvalues do not depend on 
the ultraviolet cutoff $\Lambda$. The size of the 
coupling constant $g$ that was considered small in 
Section \ref{sec:UV}, can now be understood as
actually concerning $g_\Lambda$ in the initial, 
regulated Hamiltonian that includes counterterms;
a bound state exists if $g_\Lambda$ is sufficiently 
large and this means different sizes of $g_\Lambda$ 
for different values of $\Lambda$.

Note that the finite part of the counterterm,
which amounts to the choice of $g_\lambda$ at some
value of $\lambda$, is fixed by the condition of
matching one eigenvalue with experiment. This one
condition by itself (fixing one parameter to match
one energy) would not guarantee that other
eigenvalues are cutoff independent. But it does
provide such guarantee when all matrix elements of
the $CT$ (many numbers) are calculated in the RG
procedure and fixing just one value of the
coupling constant properly determines the entire
Hamiltonian matrix so that the low energy
eigenvalues do not depend on the cutoff $\Lambda$
(and, by construction, also do not depend on the
cutoff $\lambda$). One may expect that other
matrix structures than just $\sqrt{E_m E_n}$
emerge in $H_\lambda$ with small $\lambda$, but
these structures must cease to depend on $\Lambda$ 
when $\Lambda \rightarrow \infty$ once $g_\Lambda$ 
is set according to Eq.~(\ref{af}).

In more detail, the argument that other eigenvalues of
$H_{\lambda_0}$ that are much smaller than
$\lambda_0$ will also be independent of the cutoff
$\Lambda$ once one inserts in the initial $H$ the
counterterm defined in Eq.~(\ref{CT}), is
following. Eq.~(\ref{recursion}) is valid and the
same for eigenvalues that satisfy the conditions
$|E| \ll \lambda_0$ and $1-g_0 \gg |E|/\lambda_0$.
In fact, the exact way in which all matrix
elements of $H^{(K=N-n)}(E)$ in the simple model
depend on the initial cutoff $\Lambda$ is determined 
by the dependence of $g^{(K)}$ on the sequence of 
ratios $E/\Lambda$, $bE/\Lambda$, ..., $b^{N-n}E/\Lambda
= E/\lambda$ and the initial value of the coupling
constant, i.e., after inclusion of the
counterterm, just $g_\Lambda$. For as long as
these ratios are negligible, all matrix elements of
``Hamiltonians'' $H^{(K)}(E)$ do not actually
depend on the value of $E$ and do not depend on
the cutoff $\Lambda$. Therefore, their eigenvalues
are also independent of $\Lambda$. However, one
cannot claim that all their eigenvalues are the
same because the number of eigenvalues depends on
the size of the cutoff (the smaller the cutoff the
smaller the number of eigenvalues). Only the 
eigenvalues much smaller than the smallest cutoff 
$\lambda$ in the recursion are the same and independent 
of $\Lambda$ when calculated from all Hamiltonians
$H_\lambda$ in which the coupling constant changes
from on value of $\lambda$ to another according to
the asymptotic freedom formula. 

Additional steps in evaluation of counterterms 
would be required in order to obtain corrections 
to the leading counterterm in Eq.~(\ref{CT}) due 
to the ratios $E/\Lambda_K$ differing from 0 in 
the RG recursion. Readers interested in such steps 
may consult Ref.~\cite{universality}. It is not 
necessary to discuss these additional steps here 
in order to see the problem with increasing of 
$g_\lambda$ when $\lambda$ decreases. 

However, before proceeding to the issue of large
coupling constant at low energies (in the next
section), it should be mentioned for readers used
to thinking about RG in terms of differential
equations, that the reduction of one running
cutoff on momenta to another one that is smaller,
can be made in infinitesimally small steps (in the
case of continuous momentum variables). The
corresponding changes in the Hamiltonian matrix
elements are then described by differential
equations.

\subsection{ Difficulty with $g_\lambda$ that 
             grows at small $\lambda$ }
\label{subsec:difficultywithg}

Consider Eq.~(\ref{g1model}) which implies the recursion 
\beq 
g_{\lambda/b}(E)
& = & 
g_\lambda(E) \, { 1 - E/\lambda \over 1 -
g_\lambda(E) - E/\lambda } \, .
\label{glambda}
\eeq
As discussed in the previous section, this
recursion simplifies enormously when on can 
disregard the ratio $E/\lambda$. And when one 
does, one obtains the conclusion that $g_\lambda$
increases when $\lambda$ decreases (this is
equivalent to asymptotic freedom which says
that $g_\lambda$ decreases when $\lambda$
increases). 

The increase of $g_\lambda$ toward small 
$\lambda$ causes a major difficulty because
the ratio $E/\lambda$ is compared with 
$1 - g_\lambda$ in the denominator in
Eq.~(\ref{glambda}). When the initial 
value of $g_\lambda$ at some large $\lambda$ 
is small and $g_\lambda$ increases when 
$\lambda$ decreases, at some point the effective 
coupling constant approaches 1. At this point,
the ratio $E/\lambda$ cannot be neglected
no matter how small it is. In other words,
the entire procedure can no longer be based
on simplifications that rely on smallness
of the eigenvalue in comparison to cutoff.

Not only the simplifications described in the
previous section no longer apply, but also the
entire procedure becomes unstable because the
small ratio occurs in the denominator. The small
denominator is reminiscent of small energy
denominators that occur in perturbation theory and
cause infrared singularities. However, the simple
model shows that the problem is related to the
strength of interactions that are capable of
canceling kinetic energy terms and producing
negative elements on the diagonal of the
Hamiltonian matrix. Such situation can occur 
when bound states are formed: attractive (negative) 
interactions are larger than the kinetic energies. 
In order to control what happens when $g_\lambda$ 
approaches 1, a different RG procedure seems 
required.
 
\section{ Similarity renormalization group procedure }
\label{sec:SRG}

In the standard RG procedure, one evaluates
effective Hamiltonians $H_\lambda$, with small, 
running cutoffs $\lambda$, as is illustrated 
in the simple model in the previous sections. 
One finds counterterms in the initial Hamiltonian 
with cutoff $\Lambda$ by demanding that all 
matrix elements of the effective $H_\lambda$ 
are independent of $\Lambda$ and by fixing
their finite parts by comparison with experiment
(the comparison may include conditions of
symmetry~\cite{PerryWilson}). Then the 
effective Hamiltonians can in principle be 
calculated for sufficiently small $\lambda$
to carry out computations of their eigenvalues
(or other observables) using computers.
Unfortunately, for feasibly small cutoffs, the
effective coupling constant in asymptotically 
free theories may be so large that naive 
expectations based on asymptotic freedom 
formulae may be inadequate. An alternative
approach is offered by the similarity 
renormalization group (SRG) procedure~\cite{SRG}.

In the SRG procedure, one proceeds according to
similar principles as in the standard approach
described in the previous sections. One also
completes two interrelated tasks. In the one task,
one finds counterterms. In the other task, one
evaluates effective Hamiltonians. The new idea is
that one does not ``integrate out'' any degrees of
freedom. Instead, one changes the basis states by
rotating them in the Hilbert space. The rotation
is designed in such a way that it guarantees the
new Hamiltonian $H_\lambda$ to have vanishing
matrix elements between basis states if they
differ in energy by more than $\lambda$. This
means that the matrix elements of $H_\lambda$ that
result from the second task in the SRG procedure,
are different from 0 only within an energy band of
width lambda along the diagonal. The algebra of
the procedure is designed in such a way that one
never encounters small energy denominators (the
differential version of SRG procedure has the 
same property). Moreover, the SRG procedure allows 
for direct evaluation of matrix elements of $H_\lambda$
without knowing anything about eigenvalues. The
reader will easily find all required details in
the original literature.

The SRG task of evaluating $H_\lambda$ with small
$\lambda$ aims at evaluating a near-diagonal
Hamiltonian matrix in such a way that the
calculation may be carried out in perturbation
theory with a small error that can decrease when
the order of the perturbative calculation
increases (perturbative approach is required in
realistic theories due to their complexity that
initially cannot be handled in any other way). One
may not apply perturbation theory to complete
diagonalization, because this would involve
solving also for non-perturbative features such as
bound-states. However, the SRG procedure can
produce a Hamiltonian matrix of a small width
$\lambda$ by ``rotating out'' only those
interactions that involve energy changes larger
than $\lambda$ and can be treated in perturbation
theory. The resulting $H_\lambda$ must be
diagonalized on a computer. The smaller $\lambda$,
the smaller the space of states required to find
the spectrum in the range of interest. But the
smaller $\lambda$, the higher order of
perturbative SRG evolution is required for
accuracy. A compromise must be found, and this is
not easy in QCD (see Appendix,
Section~\ref{subsec:SRGandQFT}). But there is also
a hope that already first 4 orders of the
calculation will be sufficient to identify key
operator structures in $H_\lambda$.

Of course, the SRG procedure renders the same $CT$
in the model as the one derived in
Section~\ref{subsec:CT}, Eq.~(\ref{CT}). Therefore,
the initial Hamiltonian with the huge cutoff
$\Lambda$ is given in the SRG also by
Eq.~(\ref{Hinitial}),
\beq
\label{IC}
H_{\Lambda m n} & = & 
E_m \delta_{mn} - g_\Lambda \sqrt{E_m} \sqrt{E_n}
\, .
\end{eqnarray}

The remaining discussion is focused on how one can
evaluate $H_\lambda$ using SRG equations in the
model~\cite{SRGf}. Next sections will describe the
outcome of these calculations and what this
outcome implies regarding the role of bound states
in evaluation of effective theories with
asymptotic freedom, or limit cycles. For
completeness and reader's convenience, a
perturbative SRG scheme that can be applied in
evaluation of all Poincar\'e generators in QFT and
used for the purpose of deriving dynamics of
effective constituent quarks and gluons in QCD, is
very briefly summarized in Appendix,
Section~\ref{subsec:SRGandQFT}.

\subsection{ Generators of the similarity
transformations }
\label{subsec:Generators}

The SRG procedure is used here in its differential
version. Below, prime denotes differentiation with
respect to the parameter $s = 1/\lambda^2$, chosen
for convenience. $s$ increases from 0 to $\infty$
when $\lambda$ decreases from $\infty$ to 0. 

The effective Hamiltonians are generated by the 
formula 
\beq
\label{SRGH}
H_\lambda' 
& = & 
[T_\lambda, H_\lambda ] \, , 
\eeq
and $T_\lambda$ is called the generator. The 
initial condition is set at $\lambda = \infty$
(corresponding to $s=0$), 
\beq
\label{Hs0}
H_{\infty mn}  
 & = &
H_{\Lambda m n} \, ,
\eeq
using Eq.~(\ref{IC}). Since all matrices in the 
model calculations, except for $H_0$, are functions 
of $\lambda$, the subscript $\lambda$ from now on 
will be omitted everywhere, unless it is needed
explicitly. The key point of the calculation is to 
choose the generator $T$ in a way that can shed 
some light on the increase of an effective 
coupling constant $g_\lambda$ at small $\lambda$.

One choice of $T$ originates in the beautiful flow 
equation proposed by Wegner for Hamiltonian matrices 
in condensed matter physics~\cite{Wegner1,Wegner2,Wegner3}. 
A whole range of applications of Wegner's and similar 
equations in many-particle physics is reviewed in 
Ref.~\cite{Kehrein}. $T$ in Wegner's 
equation has the form 
\beq
T_0 & = & [D, H] \, ,
\eeq
where $D$ denotes the diagonal part of the matrix $H$,
or $D_{mn} = D_m \delta_{mn}$ and 
$V_{mn} = H_{mn} \, (1- \delta_{mn})$. 
An alternative equation is obtained with the choice
\beq
T_1 & = & [H_0, H] \, ,
\eeq
which is employed in the SRG studies in nuclear
physics~\cite{PerrySzpigel,Bogner:2006srg,Bogner:2007jb}.
The calculation concerning $g_\lambda$ in asymptotically 
free theories (and limit cycle) that is reviewed
here~\cite{SRGf} is done with 
\beq
T_f & = & [G, H] \, ,
\eeq
where 
\beq
\label{G}
G & = & fH_0 + (1-f)D \, .
\eeq
For $f=0$, one has $G=D$, in which the diagonal
part of interactions is fully included, and
$T=T_0$ of condensed matter physics. For $f=1$, 
one has $G=H_0$, in which no interaction effects 
are included, and $T=T_1$ of the nuclear studies. 
For intermediate values of $f \in [0,1]$, $G$ 
includes interactions to an intermediate degree, 
correspondingly, and one can inspect what happens 
in various cases by varying $f$. 

How does Eq. (\ref{SRGH}) work? Since $T$ is a
commutator of Hermitian matrices, it is
anti-Hermitian and generates a unitary rotation of
$H$, which means that traces of all powers of
$H$ are constant. In particular, 
\beq
\left( Tr \, H^2 \right)' 
& = & 
\left( \sum_{m=M}^N \, D_m^2 \right)' 
+
\left( \sum_{m,n=M}^N \, |V_{mn}|^2 \right)' 
=0 \, .
\eeq
This means that the off-diagonal matrix elements
decrease if diagonal matrix elements increase, and
vice versa. Eq.~(\ref{SRGH}) implies
\beq 
\left( \sum_{m=M}^N \, D_m^2 \right)' 
& = &
4 \sum_{mn} D_m (G_m-G_n) \mid V_{mn} \mid^2 \; \\
&=& 2 \sum_{mn} (D_m-D_n) (G_m-G_n) \mid V_{mn} \mid^2 
\label{diagonals}
\end{eqnarray}
No negative terms appear on the right-hand-side 
when all differences $D_m-D_n$ and $G_m-G_n$ always 
satisfy the condition 
\begin{eqnarray}
(G_m-G_n)(D_m-D_n) \ge 0 \, .
\end{eqnarray}
Using notation $\Delta H_{0mn} = E_m - E_n$ and 
$\Delta H_{Imn} = H_{Imm} - H_{Inn}$, this condition 
can be rewritten, for every pair of diagonal elements 
number $m$ and $n$, as
\begin{eqnarray}
[f \Delta H_0 + (1 - f )(\Delta H_0 + \Delta H_I
)](\Delta H_0 + \Delta H_I ) \ge 0 \, .
\end{eqnarray}
Dividing by $\Delta H_0 > 0$ for $m > n$, one
obtains that if $v = \Delta H_I/\Delta H_0$ satisfies
the condition
\begin{eqnarray}
\label{vcondition}
[f + (1-f)(1 + v)] ( 1 + v) \ge 0 \, ,
\end{eqnarray}
then the sum of squares of the diagonal matrix
elements of $H$ increases and the off-diagonal
matrix elements of $H$ decrease. (By the way, in
the continuum limit for Hamiltonian matrices, such
as $b \rightarrow 1$, Eq.~(\ref{vcondition})
provides a condition on a derivative of the
diagonal matrix elements of $H_I$ with respect to
diagonal matrix elements of $H_0$.)

For $f=0$, the condition~(\ref{vcondition}) says
that $(1+v)^2 \ge 0$. This is always true and
Wegner's generator always diagonalizes Hamiltonian
matrices because the SRG evolution stops first
when all $|V_{mn}|^2$ are zero, except for
elements $V_{mn}$ for whose subscripts $D_m =
D_n$; these may in principle stay constant unless
they change due to coupling with other non-vanishing
off-diagonal matrix elements.

For $f=1$, the condition~(\ref{vcondition}) reads
$v \ge -1$. This means that the diagonal part of
the interaction must not decrease faster along the
diagonal than the free energy increases.
Convergence of $H$ to a diagonal matrix may fail
if
\beq
\frac{\Delta H_I}{\Delta H_0} < -1
\eeq
for some momenta. Since $E_m$ increases monotonically 
with $m$, the lack of convergence may occur when
$H_{Imm}$ rapidly decreases with $m$. This happens 
when a negative matrix element appears on the diagonal
among positive ones, leading to a negative eigenvalue 
that corresponds to a bound-state. The 
negative diagonal matrix elements on the diagonal 
do not guarantee that the SRG transformation stops 
driving off-diagonal matrix elements to zero, 
but it indicates that bound states may interfere
with convergence of the SRG evolution of matrices
$H_\lambda$. 

For intermediate values of $f$, two sufficient, 
mutually exclusive but not necessary conditions
for SRG evolution to bring $H$ to the diagonal 
(outside regions of degeneracy mentioned earlier), 
are
\begin{eqnarray}
\frac{ \Delta H_I}{\Delta H_0} \le \frac{1}{f-1} 
\quad \quad  {\rm or} \quad \quad 
\frac{\Delta H_I}{\Delta H_0} \ge -1 \, .
\end{eqnarray}
In the model, it happens that these conditions 
can be violated when a bound state exists. The
SRG evolution continues to bring $H$ to the
diagonal, but the effective coupling constant
may become very large. This will be explained
in the next section. 

\section{ Solutions for $H_\lambda$ }
\label{sec:Solutions}

Formula (\ref{SRGH}) produces a set of coupled 
nonlinear differential equations for all matrix 
elements of $H$, which means $(N-M+1)^2$ functions 
of $s$,
\begin{eqnarray}
D'_n & = & 2 \sum_{k=M}^N 
(G_n - G_k) V_{nk} V_{kn} \, , \\
V'_{mn} & = & - (G_m - G_n)(D_m - D_n) V_{mn}
\nonumber \\
& + & 
\sum_{k=M}^N (G_m + G_n - 2G_k) V_{mk} V_{kn} \, .
\end{eqnarray}
This is a formidable set of equations to solve.
For example, if $b=2$, one needs 37 states to 
span the energy range between $b^M \sim 1$ keV 
and $b^N \sim 70 $ TeV, and this implies 703 
functions of $s$ for real symmetric matrices 
(a complex Hermitian Hamiltonian would imply 
1369 functions). The only known way to learn 
precisely how the solutions to these equations 
look like with the initial condition at $s=0$ 
set by Eq.~(\ref{Hs0}), is to solve them 
numerically 

Fortunately, the numerical analysis (it is too
extensive to review here; readers interested in
the numerical analysis need to consult the
original literature) produces results that can be
summarized by a very simple, qualitatively
accurate analytic formula for $H_\lambda$
\beq
\label{Hlambda}
H_{\lambda m n} 
 & \sim & 
\left[ E_m \delta_{mn} - g_\lambda 
\sqrt{ E_m } \sqrt{ E_n} \right] e^{-(E_m -
E_n)^2/\lambda^2} \, ,
\eeq
which becomes exact in the limit $(E_m +
E_n)/\lambda \rightarrow 0$. This means that the
matrix elements evolve with $\lambda$ in a
coherent fashion and a relatively small number of
simple functions of $E_m$ and $E_n$ is sufficient
to reasonably well describe the evolution of all
of them. In particular, for $E_m$ and $E_n \ll
\lambda$, the exponential factor is equivalent to
1 and SRG evolution of the entire low-energy
corner of the Hamiltonian matrix is described by
just one function denoted by $g_\lambda$ in
Eq.~(\ref{Hlambda}). In analogy with the Thomson
limit in QED, this function can be extracted from
the lowest energy diagonal matrix element of the
interaction term in $H_\lambda$, $H_{\lambda MM} 
= E_M - g_\lambda E_M$. The result is that the 
following quantity is called the effective coupling 
constant in a Hamiltonian with a finite SRG cutoff 
$\lambda$,
\beq
\label{gdefinition}
g_\lambda & = & 1 -
H_{\lambda M M}/E_M \, . 
\eeq
An alternative definition of $g_\lambda$, with the
same result, could be based on the interaction
matrix elements between two basis states
corresponding to the eigenvalues $E_M$ and
$E_{M+1}$, in analogy with the example illustrated
by Eq.~(\ref{2x2}). Namely, 
\beq
\label{gdefinition2}
g_\lambda & = & - H_{\lambda M M+1}/\sqrt{E_M E_{M+1}} \, . 
\eeq
\subsection{ Asymptotically free effective 
             interactions for different generators }
\label{subsec:AFandf}

\begin{figure}[b] 
\includegraphics[scale=0.6]{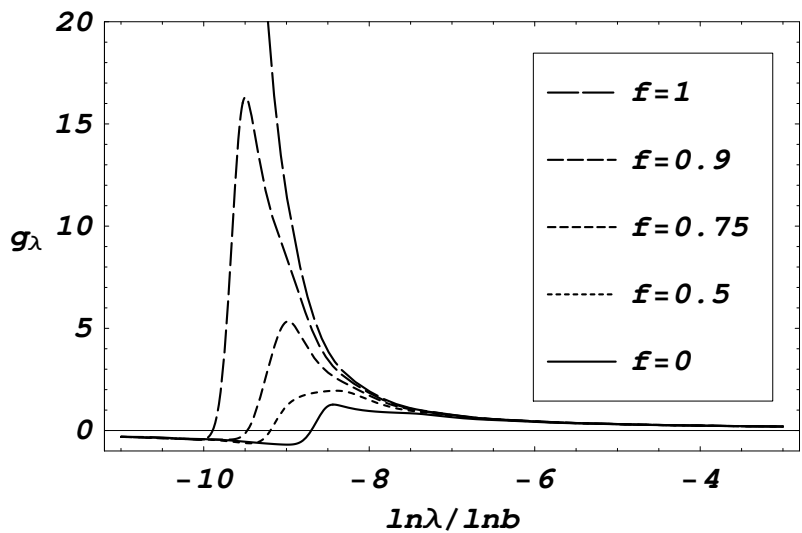} 
\caption{ \label{fig:gaf} {\small Rise
of the asymptotically free coupling constant
$g_\lambda$ at small $\lambda$, drawn as a
function of $\ln{\lambda}/ \ln{b}$ for 6 values of
$f$: $f=0$, 0.5, 0.75, 0.9, and 1 (the larger $f$,
the higher curve). The rapid increase of
$g_\lambda$ for $f=1$ below $\ln{\lambda}/ \ln{b}
\sim -8$, occurs because $\lambda$ decreases below
the scale of binding energy $E_B$. When $\lambda
\rightarrow 0$, $g_\lambda$ for $f=1$ does not
actually reach $\infty$ but $|E_B|/b^M \sim
10^{10}$. For $f=0$, the huge increase of
$g_\lambda$ is absent and instead $g_\lambda$
never exceeds order 1. See the text for
explanation.}} 
\end{figure}

The effective coupling constant defined in
Eqs.~(\ref{gdefinition}) or~(\ref{gdefinition2}), 
is a function of $\lambda$. The observation made in
Ref.~\cite{SRGf} is that the function one obtains
depends on the choice of the parameter $f$ in the
generator of the SRG transformations. A generic
example is shown in Fig.~\ref{fig:gaf}.
In this example, $b = 4$, $N=16$, $M=-25$, and
$g_\infty = g_\Lambda \sim 4/100$, where
$\Lambda = 4^{16} \sim 4 \cdot 10^9$. The
bound-state energy, $E_B \simeq 8 \cdot 10^{-6}$.

The mathematical mechanism of SRG transformations 
by which the effective coupling constant $g_\lambda$ 
increases to huge values for $f=1$, or stays near 
1 for $f=0$, or increases to intermediate values 
for intermediate values of $f$, is explained 
in detail in Ref.~\cite{SRGf}. The physical essence 
of the explanation is that SRG transformations with 
different values of $f$ place eigenvalues of 
$H_\lambda$ on the diagonal when $\lambda \rightarrow 
0$ in different places. A natural location on the 
diagonal for an eigenvalue $E$ would be a diagonal 
matrix element in the basis state $|m\rangle$ to 
which $H_0$ assigns the eigenvalue $E_m \sim E$. 

When $f=0$, interactions are fully accounted for
in the generator through $G$, see Eq.~(\ref{G}).
The reduction of $\lambda$ brings $H_\lambda$
nearer its diagonal including the interaction
energy that is responsible for binding. So, some
diagonal matrix element can naturally become
negative for some basis state $|m\rangle$ with
$E_m \sim |E| = E_B$. 

When $f=1$, interactions are completely ignored in
$G$, i.e., $G = H_0$. The diagonal matrix elements
can still approach 0 from above. But at some
point, the monotonic increase of diagonal matrix
elements along the diagonal toward large energies
(the greater subscript $m$ of a rotated basis
state the greater the diagonal matrix element) is
broken and a decrease along the diagonal (instead
of increase) occurs at some state $|m\rangle$. At
this place, the SRG transformation with $f=1$
stops reducing the size of the nearby off diagonal
matrix elements. Instead, their size increases and
maintains the strength required for reproduction
of the bound-state eigenvalue on the diagonal for
some state $|m\rangle$ with $E_m < E_B$. The shift
of the eigenvalue $-E_B$ of a fixed magnitude to a
lower-energy basis state requires a corresponding
increase of the interaction strength. For $f=1$,
the shifting of $E=-E_B$ toward low energies
continues to the very end of the SRG evolution 
at $\lambda = 0$ and places $E = -E_B$ at the state 
with the lowest allowed $E_m$ in the model: the 
bound-state eigenvalue appears as the diagonal 
matrix element $\langle M | H_\lambda | M \rangle$ 
when $\lambda \ll b^M$. As a result of this cumulative 
shift away from the natural momentum scale for the
bound-state wave function, the coupling constant 
$g_\lambda$ continues to increase in order to 
eventually produce $-E_B$ on the diagonal through 
$(1 - g_\lambda)b^M$. Thus, $g_\lambda$ increases 
toward $E_B/ b^M$, which is a huge number for a large 
negative $M$. 

When $f$ has an intermediate value between 0 and
1, the cumulative shift stops at certain state $|m
\rangle$, for which $f E_m + (1-f) (1 - g_\lambda)
E_m$ becomes negative and reproduces the
eigenvalue $-E_B$. This requires $g_\lambda$ with
$\lambda \sim E_m$ to increase only to $\sim
E_B/E_m$. 

The above scenario of how the increase of
$g_\lambda$ occurs, depending on the value of $f$,
is reflected in Fig.~\ref{fig:gaf}. Maximum of
the curve with $f = 0.75$ occurs at $-9$, instead 
of $-8.5$ in the case of $f=0$, and the maximal
value of $g_\lambda$ for $f=0.75$ is 4, or $b$ times 
larger than in the case with $f=0$. The maximum of 
the curve with $f = 0.9$ occurs at $-10$, instead of $-8.5$
in the case of $f=0$, and the maximal $g_\lambda$
for $f=0.9$ is 16, or $b^2$ times larger than in
the case with $f=0$, and so on. The case with
$f=1$ leads to apparently indefinite and
accelerating increase that smoothly continues the
inverse logarithmic growth that characterizes
asymptotic freedom above the scale of binding. 
The transition from an asymptotically free RG 
behavior to the behavior dominated by bound 
states occurs when $g_\lambda$ crosses 1, cf. 
Section~\ref{subsec:difficultywithg}.

In summary, the model demonstrates a possibility
that an unlimited increase in the asymptotically
free interaction at small energies is caused by
omission of interactions responsible for the
existence of a bound state in the generator of the
SRG transformations. When the generator fully
accounts for the interactions responsible for
binding, the magnitude of the coupling constant
never significantly exceeds 1. 

The model example is potentially important because
it suggests that the increase of the coupling
constant in asymptotically free theories that is
observed in perturbation theory (corresponding to
$G=H_0$ and $f=1$) may be not related to the
phenomenon of confinement but to the phenomenon of
formation of bound states. The latter is in
principle a simpler one to handle in theory than
the former. If it were indeed the case,
application of the SRG procedure to QCD may help
theorists in penetrating the range of momentum
scales near $\Lambda_{QCD}$ and explaining
hadronic states without immediate necessity to
solve the problem of confinement that is relevant
at distances much larger than the size of
individual hadrons (see also Appendix,
Section~\ref{subsec:SRGandQFT}).

\subsection{ Comment on asymptotic freedom and limit cycle } 
\label{subsec:AFandLC}

An apparently very small alteration of the model 
introduced in Section~\ref{sec:Model} leads to 
new ways of thinking about asymptotic freedom
in terms of limit cycle. RG limit cycles were 
discovered in the context of strong interactions 
in Ref.~\cite{WilsonGML}, using the nowadays 
standard RG procedure. The SRG approach to limit 
cycle is based on Ref.~\cite{SDGeffective}. The 
initial Hamiltonian matrix of the altered model 
is 
\begin{eqnarray}
\label{HinitialLC}
H_{\Lambda mn} & = & \sqrt{ E_m E_n} \, 
\left[ \, \delta_{mn} - g_\Lambda - i h \, {\rm sgn}(m-n) \, \right] \, . 
\end{eqnarray}
where $i = \sqrt{-1}$ and the new coupling
constant $h$ can be an arbitrarily small real
number. If $h$ is an arbitrary irrational number,
the model typically exhibits chaotic RG behavior.
When $h = \tan{\pi\over p}$ with $p$ an integer
greater than 2, a limit cycle occurs. Namely,
repeating the same procedure as described in the
previous sections for $h=0$, one finds that the
coupling constant $h$ does not evolve with
$\lambda$ at all, while the coupling constant
$g_\lambda$ that replaces $g_\Lambda$ in a similar 
way as $g_\lambda$ replaced $g_\Lambda$ for $h=0$,
oscillates with $\lambda$ with a multiplicative
period $b^p$. This means that $g_{\lambda_1}$ has
the same value as $g_{\lambda_2}$ if $\lambda_1 =
(b^p)^k \lambda_2$ with an arbitrary integer $k$.
The periodicity (cycle) of the coupling constant
is associated with existence of infinitely many
bound states (in the limit $M \rightarrow -
\infty$) whose binding energies form a geometric
series converging on zero with quotient $1/b^p$.

RG evolution of an asymptotically free interaction
forms a part of a limit cycle for all values of
$\lambda$ for which $h$ is very small in
comparison to $g_\lambda$. In this range, $h$ does
not matter and the Hamiltonians of the model
defined by Eq.~(\ref{HinitialLC}) evolve in the
same way as in the case of Eq.~(\ref{IC}) (see
Fig.~3 in Ref.~\cite{SDGeffective}). Consider the
bottom-up evolution in which $\lambda$ grows and
the coupling constant $g_\lambda$ decreases as an
inverse of a logarithm of $\lambda$ until it
becomes comparable with $h$. Then, instead of
$g_\lambda$ continuing its logarithmic falloff
indefinitely, $h$ takes over and forces
$g_\lambda$ to change sign and subsequently grow
in size. This continues until a new high-energy
bound state is reconstructed. When this happens,
$g_\lambda$ switches sign to positive again and
grows rapidly above 1, its magnitude depending on
the size of $f$. The switch occurs at the scale
$\lambda$ on the order of the energy (momentum)
where the new bound-state energy is located on the
diagonal. Further increase of $\lambda$ produces a
falloff like in asymptotic freedom again, until
$g_\lambda$ decreases again down to the size of
$h$.

The range of scales for which the cycle looks like
asymptotic freedom is given by the period of the
cycle, characterized by the factor $e^{\pi/h}$,
which can be very large when $h$ is very small.
Such behavior of the simple model is of general
interest because it suggests that the hierarchy
problem may stem from continuing perturbative
analysis for coupling constants $g_\lambda \ll 1$
while overlooking formation of new generation of
bound states due to very small, and so far
unknown, interactions of the type represented by
the coupling constant $h$. But in order to see
their presence, one has to use SRG procedure with
the generator that includes interactions in $G$.

\section{ Conclusion} 
\label{sec:C}

The simple model study shows that the SRG
procedure may be a suitable tool to handle the
increase of the coupling constant $g_\lambda$ in
QCD when $\lambda \rightarrow \Lambda_{QCD}$. If
the generator of SRG transformations does not
include interactions in $G$, $f=1$ in
Eq.~(\ref{G}), the effective coupling constant
increases to very large values quickly as soon as
the SRG scale parameter $\lambda$ becomes
comparable with the momentum scale that
characterizes formation of a bound state. In QCD,
the corresponding momentum scale would be much
larger than the scale associated with confinement
because the size of a single hadron is much
smaller than the distances at which confinement
matters. If the generator of SRG transformations
includes interactions in $G$, the SRG parameter
$\lambda$ can be brought down to the momentum
scale that characterizes bound states and the
coupling constant does not increase to large
values. These model results suggest that the SRG
procedure should be applied to QCD because it may
offer help in understanding the binding mechanism
for quarks and gluons using expansion of
$H_{\lambda QCD}$ in powers of an effective
coupling constant and without need for prior
understanding of confinement. Interestingly enough, 
the SRG procedure may also help us establish a 
connection between asymptotic freedom and limit 
cycle. In this respect, the model shows that in 
order to handle the case of limit cycle the 
generator of SRG transformations must include 
interactions in $G$.

\vskip.2in
\centerline{\bf Acknowledgement}
\vskip.1in
The author would like to thank Robert Perry of the 
Ohio State University for many discussions concerning 
RG procedures and hospitality extended to the author 
during his visits to OSU. It is also the author's 
pleasure to thank Micha{\l} Prasza{\l}owicz and his 
colleagues at the Jagiellonian University for organizing 
another excellent Cracow School of Theoretical Physics, 
and for the outstanding hospitality they graciously 
provide.

\section{Appendix}
\label{sec:A}

The Appendix describes examples of QFT counterparts 
of the concepts concerning RG procedure and bound 
states that are introduced in the main text in a simple 
model. 

\subsection{ Potentials with $\delta$-functions }
\label{subsec:delta}

When one calculates corrections to the Coulomb
potential in QED, one obtains a $\delta$-function
as the Uehling term in $H_I$. This term contributes 
a small part in the Lamb shift. The Coulomb
potential, $- \alpha/r$, in the Schr\"odinger
equation is changed to~\cite{Peskin-delta} 
\beq
V(\vec r \,) & = & - {\alpha \over r } 
                   - {4 \alpha^2 \over 14 m_e^2 } \, 
\delta^{(3)}(\vec r \,) 
\eeq 
The correction appears suppressed by
$\alpha \sim 1/137$ in comparison to the Coulomb
potential. In first-order perturbation theory, one
obtains truly tiny corrections ($\sim 10^{-7}$ eV
for $2S$ states~\cite{Peskin-delta}). However, the
second order correction involves multiplication of
the $\delta$-function by itself and produces
infinity. The problem is not merely due to the use
of perturbation theory instead of solving the
Schr\"odinger equation exactly since the
$\delta$-function (or a similarly singular
function) leads to so strong a potential that the
wave function collapses unless the singular
function is somehow replaced by a regular one. One
can attempt to derive a regular expression in
perturbation theory, say, by limiting the range of
momenta in intermediate states from above by the 
electron mass $m_e$ times the speed of light. On the
other hand, a complete analysis should include the
formation of a bound state, and bound states are
not describable in perturbation theory. Some form
of an effective theory is necessary~\cite{Pachucki} 
(see also Ref.~\cite{Weinberg}). In contrast with 
other approaches, the singular potentials that 
are obtained in QCD (or QED) in the SRG approach, 
are always effectively regulated by form factors 
of width $\lambda$, the width playing the role of 
a renormalization group parameter (e.g., see 
Ref.~\cite{QQ}).

Studies of $\delta$-function potentials by
physicists have a long
history~\cite{Salpeter,Lee,Yamaguchi,Heisenberg,
Zeldovich,BerezinFaddeev,Jackiw,Jaffe,Derezinski}
and there exist mathematical textbooks on the
subject~\cite{Yafaev,AlbeverioKurasov}. The simple
model used in this lecture can also be derived by
discretizing momentum on a logarithmic scale in
the $S$-wave Shr\"odinger equation for a particle
moving on a plane in the presence of a potential
proportional to a $\delta$-function (see the
original literature).

\subsection{ SRG procedure in QFT }
\label{subsec:SRGandQFT}

Since the model discussed in this lecture appears
deceptively simple, this section provides a
telegraphic overview of how the SRG procedure
applies in QFT. The original SRG application to
QCD was outlined in Ref.~\cite{LFQCD}, for
calculating matrix elements of light-front (LF)
effective $H_{\lambda QCD}$ using perturbation
theory. A perturbative calculus for $H_{\lambda
QCD}$ with $\lambda \gg \Lambda_{QCD}$ in terms of
creation and annihilation operators, was developed
in Ref.~\cite{RGPEP}, and shown in
Ref.~\cite{algebra} to be able to produce in
simple scalar theories not only the Hamiltonian
operator (the energy-momentum tensor component
$T_\lambda^{+-}$) but also other generators of the
Poincar\'e algebra. This calculus produced LF QCD
Hamiltonians with running coupling constant
$g_\lambda$ and recently led to a reasonable
description of heavy-quarkonium spectra, still
using crude and as yet unverified approximations
concerning terms order $g_\lambda^4$~\cite{QQ}.
The calculus is invariant with respect to 7
independent Poincar\'e symmetries (including 3
boosts), satisfies required cluster
property~\cite{Weinberg}, and guarantees that the
resulting Hamiltonians $H_\lambda$ have form
factors in the interaction vertices of width
$\lambda$ in energy, the width being the SRG
parameter. All these features are required when
one attempts to derive the parton model and
spectroscopy of hadrons from one and the same
Hamiltonian formulation of QCD.

The point of departure is a canonical LF Hamiltonian
in QFT, which requires regularization and counterterms,  
\begin{eqnarray}
\label{HQCDinitial}
H = \left[ H_{can} + H_{CT} \right]_{reg}  \, .
\end{eqnarray}
The canonical gluon field $A$ in $A^+=0$ gauge 
(LF in the Minkowski space-time is defined by the 
condition $x^+ = x^0 + x^3 = 0$ and $A^+=0$ 
means that $A^0 + A^3 = 0$) and quark field $\psi$, 
are expanded into their Fourier components at
$x^+=0$. The Fourier components correspond to 
canonical creation and annihilations operators 
for bare quarks and gluons (or other bare particles 
in other theories than QCD). These canonical 
operators, say $q_{can}$, are related by a 
unitary transformation $U_\lambda$ to their 
counterparts for effective particles, say $q_\lambda$,
\beq
q_\lambda & = & U_\lambda \, q_{can} \,
U_\lambda^\dagger  \, ,
\eeq
and $U_\infty = 1$. The Hamiltonian $H = H(q_{can})$ 
is assumed equal to $H_\lambda(q_\lambda)$, but 
the coefficients in expansion of $H_\lambda(q_\lambda)$ 
in powers of $q_\lambda$ are different from the coefficients
of expansion of $H(q_{can})$ in powers of $q_{can}$. 
The object of the SRG calculation are the
coefficients. One actually works in the constant 
basis in the operator space and evaluates $
{\cal H}_\lambda = H_\lambda(q_{can})$. The SRG 
equation is
\beq
{d \over d\lambda} {\cal H}_\lambda & = & [{\cal T}_\lambda,
{\cal H}_\lambda ]  \, ,
\eeq
where
\beq
{\cal T}_\lambda & = & {\cal U}'_\lambda {\cal
U}^\dagger_\lambda \, ,
\eeq
and ${\cal U}_\lambda$ corresponds to $U_\lambda$. 
Assuming the initial condition of
Eq.~(\ref{HQCDinitial}),
\beq
{\cal H}_\infty & = & \left[ H_{can} + H_{CT}
\right]_{reg} \, ,
\eeq
one derives $H_\lambda$ from the formula
\beq
{\cal H}_\lambda = {\cal H}_\infty + \int_\infty^\lambda d\omega
[{\cal T}_\omega , {\cal H}_\omega ] 
\end{eqnarray}
order by order in perturbation theory, eventually
replacing $q_{can}$ by $q_\lambda$.

The key is ${\cal T}_\lambda$. ${\cal
H}_\lambda$ is split into a bilinear term in
$q_{can}$, denoted by ${\cal H}_1$, and the
remaining terms, denoted by ${\cal H}_2$. 
${\cal H}_2$ is assumed to contain the form factors
$f_\lambda$ in vertices (the letter $f$ used here 
has nothing to do with the letter $f$ used in $G$ 
in Section~\ref{subsec:Generators}). Thus,
${\cal H}_2 =  f {\cal G}_2$, and the form factor
is defined for all operators in the same way
as in the following example:
\beq
{\cal G} 
& = & 
\int [123] \, g(1,2,3) \, 
q^\dagger_{can 1} q^\dagger_{can 2}
q_{can 3} \, , \\
f {\cal G} 
& = & 
\int [123] \, f(123) \, g(1,2,3) \, 
q^\dagger_{can 1} q^\dagger_{can 2}
a_{can 3} \, , \\
\label{fl}
f(123) 
& = & 
\exp{\left[ - ({\cal M}_{12}^2 - {\cal
M}_3^2)^2/\lambda^4 \right]} \, .
\eeq
Namely, when an operator $\cal G$ contains a product 
of $m$ creation and $n$ annihilation operators, 
the form factor operation inserts the same function 
$f$ of the difference between invariant 
masses squared, ${\cal M}_m^2 = \left( \sum_{i=1}^m p_i
\right)^2$ and ${\cal M}_n^2 = \left( \sum_{j=1}^n k_j
\right)^2$, that are associated with the particle momenta
using eigenvalues of ${\cal G}_1$.  
The SRG generator in QFT is defined by the
commutator~\cite{RGPEP}
\beq
\left[{\cal T}, {\cal G}_1 \right]
& = & 
\left[ (1-f){\cal G}_2 \right]' \, .
\eeq
This choice guarantees that the resulting
interactions are connected (cluster property)
and perturbation theory for evaluation of 
${\cal H}_\lambda$ never contains small
energy denominators, provided the function $1-f$ 
vanishes faster than linearly in the difference
of invariant masses (for the function adopted in
Eq.~(\ref{fl}), $1-f$ vanishes quadratically).  

\subsection{ Questions concerning
             SRG and AdS/CFT in QCD }
\label{subsec:AdSCFT}

A few questions are posed here regarding the
possibility that the SRG procedure in QCD and 
a RG interpretation of AdS/CFT duality in the 
context of QCD, can be related. These questions 
reflect how little is understood about the SRG 
procedure and formation of bound states in QCD.

SRG procedure is expected to render Hamiltonians
$H_\lambda$ whose eigenstates in the case of LF
QCD (as a part of the standard model) can be used
to define hadrons and calculate hadronic
observables. A hadron eigenstate is a linear
combination of Fock components $|n\rangle$ with
various values of $n$, where $n$ denotes the
number of operators $q^\dagger_\lambda$ in a
product that is used to create the state
$|n\rangle$ from the LF vacuum state $|0\rangle$.
The corresponding wave functions $\psi_{n \lambda}$ 
will depend on $\lambda$ while the eigenvalue
and the eigenstate as a whole will not. Expansion
of a hadron state into components with various
numbers of effective constituents will be very
broad in energy for $\lambda \gg \Lambda_{QCD}$
(corresponding to a parton model state for probes 
acting with large energy transfers) and may be 
dominated by the lowest constituent-number components 
for $\lambda \sim \Lambda_{QCD}$ (corresponding to 
the constituent model for probes acting with small 
energy transfers). 

The family of Hamiltonians $H_\lambda$, all
equivalent, can be imagined to be related to a
family of complex effective classical Lagrangian 
densities ${\cal L}(\lambda, x)$ in the Minkowski
space-time, all of them representing the same
theory. The parameter $\lambda$ can be considered 
a 5th dimension in a larger theory and one can 
contemplate actions of the form 
\beq
S & = & \int d\lambda d^4 x \, {\cal L}(\lambda,x)
\, .
\eeq
In such action, the coupling constant $g_\lambda$
can be considered a function of the 5-th
coordinate $\lambda$, and also a function of $x$.
Eventually, no hadronic observable will depend on
$\lambda$. Nevertheless, a rich dynamics could be
considered in the 5-dimensional space, or in more
dimensions, in which the family of equivalent
Lagrangians labeled by $\lambda$ could emerge as 
describing configurations that dominate appropriate 
path integrals. 

Suppose the dependence on $\lambda$ is analogous
to Polyakov's RG interpretation of the 5th
(Liouville) dimension in string
theory~\cite{PolyakovWall,stringsQCD}, 
cf.~\cite{MaldacenaReport}.
In the SRG procedure, however, one always deals
with effective quantum field theories in the
Minkowski space. Can these effective theories
explicitly include scale dependent gluonic string
dynamics that could be ``a shadow on the wall'' of
what happens in the bulk? According to~\cite{QQ},
one may expect explicit formation of quantum
gluonic strings in LF Hamiltonian QCD in the
Minkowski space when one considers eigenstates in
which relative motion of constituents corresponds
to considerably larger energies than
$\Lambda_{QCD}$ and components with $n \gg 1$ for
effective gluons are important. Could they
correspond in some way to a string theory in the
bulk?

There is another line of thinking that leads to
similar questions, based on the observation that
the LF wave functions $\psi_{n \lambda}$ can be
used to calculate hadronic form factors using
formulae similar to the ones considered by Brodsky
and de Teramond in
Refs.~\cite{BrodskyTeramond2,BrodskyTeramond1},
with additional factors that will result from
using the SRG procedure. Can one connect the
holography proposed by Brodsky and de Teramond
with the $\lambda$-dependent hadronic constituent
distributions (densities obtained using sums of
moduli squared of the LF wave functions $\psi_{n
\lambda}$), additional SRG factors, and
off-energy-shell old-fashioned scattering
amplitudes for effective constituents with form
factors of width $\lambda$, all of which would
result from $H_{\lambda QCD}$ obtained from the
SRG procedure?

Such questions will not be easily answered. For
example, there is a potential complication
involved in QCD due to an infrared limit
cycle~\cite{Braaten}. On the other hand, the model
described here shows that the SRG procedure can
deal also with limit cycles by including
interactions in the generator.

\end{document}